\documentclass[10pt,letterpaper]{article}
\usepackage{ol2}
\usepackage{varioref}
\usepackage{graphicx}
\usepackage[T1]{fontenc}
\usepackage{subfigure}
\usepackage{cite}
\usepackage{amsmath}
\usepackage{amsfonts}
\usepackage{stmaryrd}
\usepackage{url}

\begin{document}

\title{Miniaturized fiber dosimeter of medical radiations on a narrow optical fiber}

\author{Miguel Angel Suarez$^1$, Tony Lim$^2$, Louise Robillot$^1$, Valentin Maillot$^1$, Thomas Lihoreau$^3$, Patrick Bontemps$^2$, Lionel Pazart$^3$, Thierry Grosjean$^1$}

\address{$^1$ FEMTO-ST Institute UMR 6174, Univ. Bourgogne Franche-Comte --CNRS -- Besancon, France}

\address{$^2$Service d'Oncologie-radiotherapie, CHU Besancon, F-25000 Besancon, France }

\address{$^3$Inserm CIC 1431, CHU Besancon, F-25000 Besancon, France}

\email{thierry.grosjean@univ-fcomte.fr}

\begin{abstract}
Fiber dosimeters have recently drawn much interest for measuring \textit{in vivo} and in real time the dose of medical radiations. This paper presents the first miniaturized fiber dosimeter integrated at the end of a narrow 125 $\mu$m outer diameter optical fiber. Miniaturization is rendered possible by exploiting the concept of a leaky wave optical antenna for interfacing the scintillators and the fiber, and by taking advantage of the low propagation loss of narrow silica fibers and high detection yield of single-pixel photon counters.  Upon irradiation at 6 MV, our fiber probe led to a linear detection response with a signal-to-noise ratio as high as 190 in air. Although implemented with inorganic scintillators and fiber, our fiber probe induces an intensity contrast in the impinging radiation lower than 0.9\% over an area of 0.153 mm$^2$. Our nano-optically driven approach opens the route for ultra-compact fiber dosimeters of negligible footprint in the radiotherapeutic processes, even with non-water equivalent fibers and scintillators. A large panel of therapies relying on ionizing radiations (photons or charged particles) may take benefit of this nano-optically-driven technology.
\end{abstract}

\maketitle

\newpage

\section{Introduction}

Recent advances in radiation therapies have prompted the need for tools to accurately probe ionizing radiations at small scales, in confined environments, in real time and with compact invisible devices. Among possible technological strategies, the combination of scintillators and optical fibers has been foreseen to provide end-users with highly flexible dosimeters capable of a real-time \textit{in vivo} X-ray monitoring \cite{beddar:pmb92a,beddar:pmb92b,wells:ijrobp94,letourneau:mp99,beddar:ieee01,archambault:mp06,beddar:rm06,pittet:saa09,moon:ari12,mccarthy:ieee14,stanton:nanoscale14,belley:mp15,zhuang:ox16,deandreas:ieee17,archer:sr17}. In practice though, the low coupling efficiency between scintillators (also called phosphors) and optical fibers has thus far limited their experimental realization onto broad multimode optical fibers (of near-millimeter diameters), which represents limits for the technology. For instance, exigencies of probe invisibility with regards to the therapeutic process has imposed the use of water-equivalent plastic scintillators and fibers \cite{beddar:pmb92a,beddar:pmb92b,wells:ijrobp94,letourneau:mp99,beddar:ieee01,archambault:mp06,beddar:rm06,moon:ari12,archer:sr17,frelin:mp05,justus:ao04,clift:pmb02}. However, such probes suffer from degraded signal-to-noise ratio due to the existence of a spurious Cerenkov effect within the fiber \cite{lee:nimprs07,law:ol07}. Filtering out the Cerenkov signal is possible at the expense of a critical broadening of the probe in the context of intracorporeal dosimetry (paired-fiber configuration) \cite{wells:ijrobp94,lee:nimprs07,yoo:ox13}. 

Scaling down this on-fiber technology by approximately one order of magnitude, to make it integrated at the end of a narrow optical fiber of about the size of a human hair (80-125 $\mu$m),  would provide new opportunities in the detection of ionizing radiations. In this paper, we demonstrated the first miniaturized fiber dosimeter integrated at the end of a narrow 125 $\mu$m outer diameter optical fiber. The drop of detection efficiency accompanying the shrinkage of both the fiber and scintillating cell is compensated by implementing a key-interface in-between the luminescent material and the fiber. Based on the concept of optical antenna \cite{novotny:book}, such an photonic interconnection enables redirecting the radiation-triggered luminescence towards the fiber and efficiently phase-match the optical waves to the fiber modes.  

Optical antennas  have demonstrated performances in controlling the emission directionality of fluorescent optical sources \cite{novotny:natphot11}. Recently, the horn nano-optical antenna has been proposed to transfer the emission of a point-like dipole source into an optical fiber \cite{grosjean:ox13}.  This antenna has then been successfully used to interface a scintillating cluster to a single-mode optical fiber, thereby leading to a micron size X-ray fiber probe \cite{xie:ol17}. Such a system however finds limits in probing the high energy radiations used in medical therapies. Here, we overcome limits using the concept of leaky-wave optical antenna \cite{balanis:book,wang:ox11,peter:nl17,monticone:ieee15,liu:prb10} in a reciprocal optical approach.   

Beyond gain of compactness, the probe miniaturization opportunity offered by such a nano-optically driven technology leads to near-transparent fiber detectors even with inorganic non water equivalent fibers and scintillators \cite{shionoya:book,weber:jl02,eijk:pmb02}.  The use of narrow fibers offers other advantages such as lower propagation loss than plastic fibers and the use of mono-pixel photon counters which develop higher detection yield than the multi-pixel systems used with plastic fibers.

\section{Principle}

The fiber-mediated coupling process of luminescent material to a photodetector can be accurately described using reciprocity. If the detector is replaced by a light source at $\lambda$, the intensity distribution at the output surface of the fiber in contact to the scintillators provides a reciprocal map of the coupling efficiency of the scintillators to the detector through the fiber. The regions of higher leakage reciprocally reveal the regions of higher detection efficiency. In this respect, any source of leakage at the fiber end is reciprocally converted in a coupling region of an incoming light into the fiber. Therefore controlling the leakage at the end portion of a fiber reciprocally enables optimizing the detection sensitivity of a fiber dosimeter based on that fiber. 

\begin{figure}[htbp]
\centering\includegraphics[scale=0.55]{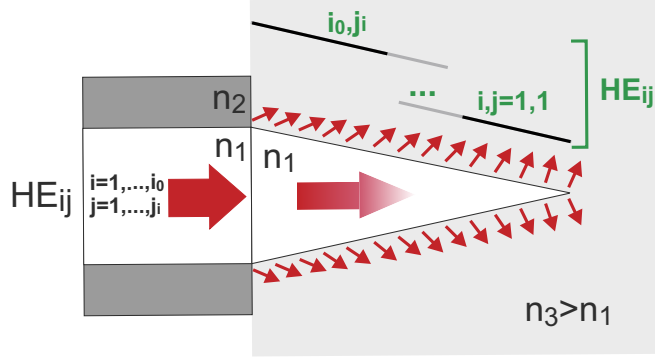}
\caption{Principle of our fiber probe: reciprocal approach. Our leaky-wave optical antenna is constituted of a cone surrounded by a high refractive index material. The structure is positioned at the end of the fiber. Light injected into the fiber is turned into a leaky wave within the optical antenna (total internal reflection is deliberately lost at the cone interface). Each incoming fiber mode leads to a leakage distribution along the cone surface defined by its dispersion properties. While increasing the orders of the fiber modes, the leakage occurs from the apex and move towards the basis of the cone. By properly designing such a leaky-wave optical antenna, the multimode nature of the fiber ensures an homogeneous leakage all along the cone, even if the structure is sharp and elongated. Reciprocally, the coupling area between a high permittivity luminescent material and the fiber is dramatically enhanced as compared to a cleaved fiber, thus leading to improved detection efficiencies.}
\end{figure}

Leaky-wave antennas belong to the family of non-resonant travelling-wave antennas \cite{balanis:book}. They achieve broadband directional radiation in free space from a lossy waveguide.  This concept has been recently extended to optics for achieving controlled emission from a tiny fluorescent source \cite{peter:nl17}. The loss channel was here provided by a high refractive index substrate in a "`fast-wave"' process. The mode within the cone continuously loses energy upon propagation, which is directionally released in the substrate, by virtue of momentum matching \cite{balanis:book}. Importantly, such a leakage mechanism is tailored by the incoming fiber mode itself via its intrinsic dispersion properties.  

\begin{figure}[htbp]
\centering\includegraphics[scale=0.95]{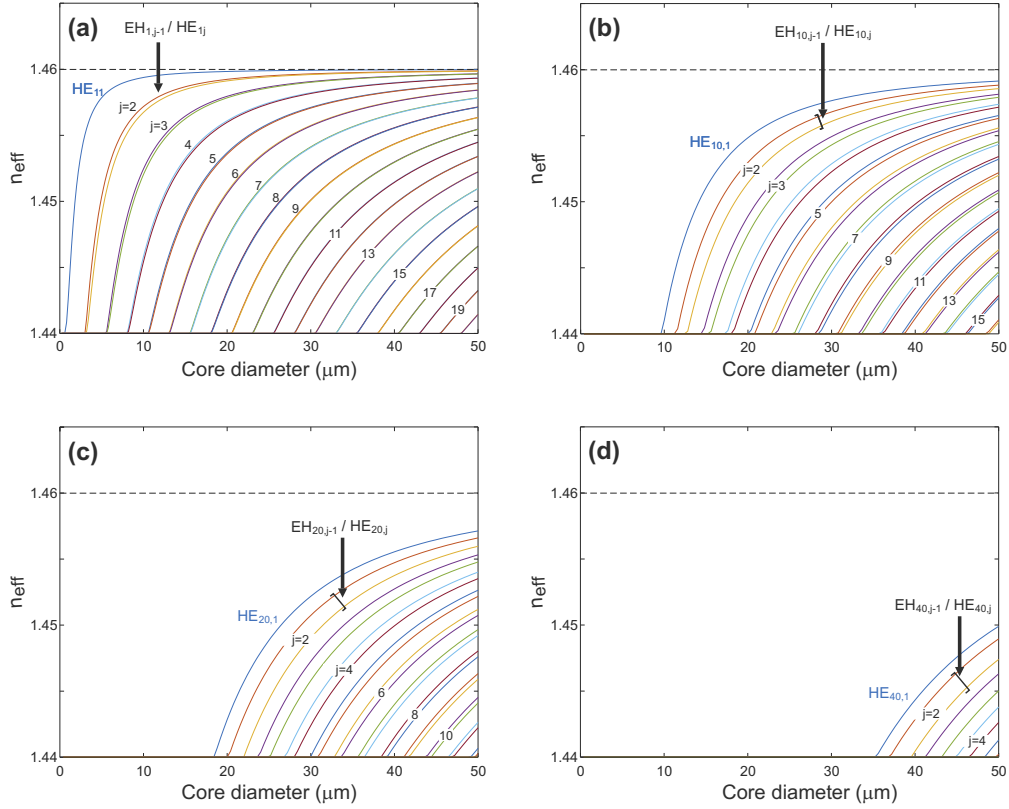}
\caption{Dispersion properties of a multimode fiber with core and cladding indices equal to 1.46 and 1.44, respectively: effective refractive index of the $HE_{ij}$ and $EH_{ij}$ fiber modes ($i$ and $j$ are integers) as a function of the core diameter. (a) $i=1$, (b) $i=10$, (c) $i=20$ and (d) $i=40$.}
\end{figure}

Our concept of a leaky-wave optical antenna is shown in Fig. 1. It consists of a dielectric cone on top of a cleaved optical fiber, surrounded by a high permittivity medium (used as leakage medium). The cone permittivity and base diameter match those of the fiber core.  Multimode fibers are considered, with core diameters ranging from 25 $\mu$m to 100 $\mu$m.  Such fibers are known to propagate four types of modes, namely the $HE_{ij}$, $EH_{ij}$, $TM_{0j}$ and $TE_{0j}$ modes ($i$ and $j$ are integers) \cite{marcuse:book}. Figure 2 shows the effective indices of the $HE_{ij}$ and $EH_{ij}$  modes as a function of the core diameter, for core and cladding refractive indices of 1.46 and 1.44, respectively. We see that these modes reach cutoff at higher core diameters for increasing values of $i$ and $j$. According to that property, the fiber modes of higher orders (i.e., of lower refractive indices and larger cutoff diameters) will undergo maximum leakage immediately after entering the cone whereas the modes of lower orders (i.e., of higher refractive indices and smaller cutoff diameters) will show the higher leakage close to the cone apex.

We numerically studied this antenna concept using finite difference time domain method (FDTD) \cite{taflove:book}.  We focused on a 50-micron core diameter multimode fiber with core and cladding refractive indices of 1.46 and 1.44, respectively. The cone length is chosen to be equal to 250 $\mu$m and the surrounding medium has a refractive index of 2.3. Fig. 3(a) represents the leakage accumulated along the cone for impinging $HE_{11}$ and $HE_{47,3}$ modes, respectively.  We see that $HE_{47,3}$ show maximum leakage in the first half of the cone whereas the power of $HE_{11}$ leaves the cone in the last 70 $\mu$m of the structure, close to its apex. The large number of fiber modes in between these two extrema ensure a near-homogeneous leakage all along the cone, as shown Fig. 3(b).

Reciprocally, part of the light isotropically generated within the highly refractive medium will be incoupled into the fiber in a mirror mechanism of the above-described mode leakage. As a result, the overall surface of the elongated cone will participate with near constant efficiency to the in-fiber light collection process. The light collection area of the fiber is thus extended by a factor of $h/R$ regarding the cleaved fiber, where h is the cone height and R is the core radius. In the configuration under consideration, the collection area is enhanced by a factor of 10. The fiber modes with the lower (higher, respectively) $i$ and $j$ indices will propagate the light mainly collected at the cone apex (at the cone basis, respectively). 

\begin{figure}[htbp]
\centering\includegraphics[scale=1.12]{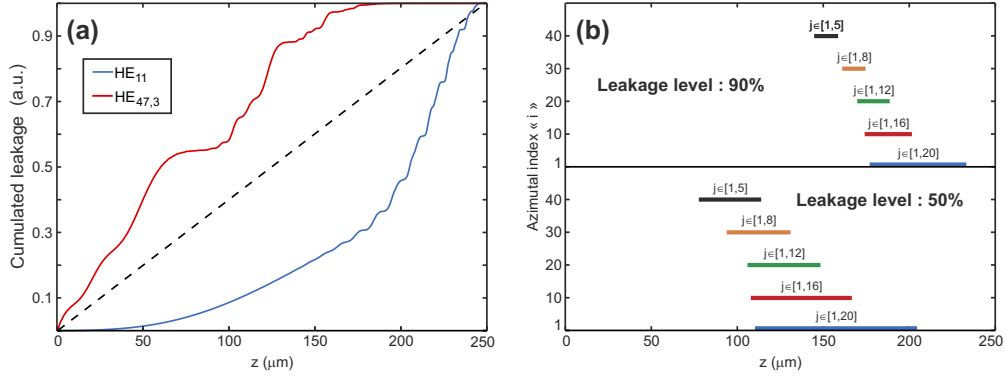}
\caption{(a) Simulation by FDTD of the accumulated leakage (in fraction of the incoming power) along the cone axis for the $HE_{11}$ and $HE_{47,3}$ modes. The dashed line represents constant leakage along the cone.  (b) Positions along the cone axis $(0z)$ where 50 \% (upper panel) and 90 \% (downer panel) of the power of the incoming fiber modes $HE_{ij}$ has leaved out the cone, for $i$=1, 10, 20, 30 and 40.}
\end{figure}

The concept of a miniaturized fiber dosimeter emerges from considering the luminescent high refractive index medium of the antenna as being a scintillating material under irradiation (i.e., a high density of luminescent point-like sources embedded in a semiconductor matrix).  Given the scattering phenomenon within scintillation materials, we limited our analysis of the optical leakage/collection efficiency at the cone surface.

\section{Demonstration}

\subsection{Fabrication}

The miniaturized fiber dosimeters have been engineered onto narrow 125-micron outer diameter multimode fiber (core diameter of 100 $\mu$m). The fiber end facets were first etched in a buffered fluorhydric acid to obtain elongated cones. The cones were about 500 $\mu$m long (see Fig. 4(a)), leading to a cone surface about 10 fold larger than the surface of the fiber core. Next, scintillating powder was mixed with photosensitive emulsion and attached to the fiber cone by a photopolymerization involving an illumination via the fiber (i.e., light is injected from the other end facet of the fiber). We chose an inorganic scintillating material, an europium-doped gadolinium oxysulfide (Gd$_2$O$_2$S:Eu) from Phosphor Technology, which generates visible light upon exposure to ionizing radiations (photons or charged particles). Such type of scintillators demonstrated good performances and linearity in the local probing of ionizing radiations in fibered architectures \cite{zhuang:ox16,mccarthy:ieee14}. Figure 4(b) displays the optical image of a resulting miniaturized fiber probe. The scintillating material takes the form of a quasi-uniform 85-micron thick layer that selectively covers the cone. The maximum width of the probe does not exceed 250 $\mu$m, i.e., the full diameter of the optical fiber with its plastic cladding. As an option to increase detection efficiency, the probe can be metal coated with a thin aluminum layer (200 nm thick) aimed at back-reflecting the initially wasted light leaving the probe outside the fiber. Aluminum is known for its good transparency to ionizing radiations and high reflectivity to light. Radiation attenuation of such thin layers is negligible \cite{nist2}. 

\begin{figure}[htbp]
\centering\includegraphics[scale=1.19]{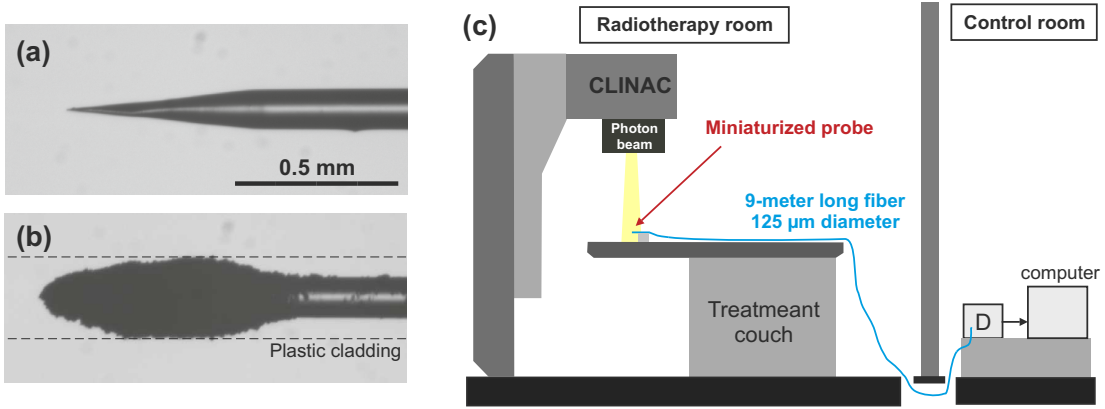}
\caption{(a) Microscope image of the tapered fiber. (b) Microscope image of the fiber probe obtained after the scintillator-to-taper attachment. (c) Schematics of the experimental set-up.}
\end{figure}

\subsection{Experimental set-up}

The miniaturized probes were engineered at one of the two ends of 9-meter long fibers covered with 0.9 mm black hytrel cladding, the other end being terminated with a FC-PC connector.  Their detection performances were tested under a clinical linear accelerometer used for external radiation beam therapy (Varian; Fig. 4(c)). The fiber probe was positioned on the treatment bed at the center of the ionizing beam with a field size of $ 5 \times 5$ cm$^2$ at a SSD (Source to Surface Distance) of 100 cm.  Upon irradiation, europium-doped gadolinium oxysulfide shows an emission spectrum peaking at 627 nm. At such a wavelength, propagation losses of the fiber (Thorlabs) are lower than 10.1 dB/km.  The optical signal leaving the fiber was recorded with an Aurea Technology SPD-A-VIS single pixel photon counter located in the control room. This optical detector ensures a quantum yield larger than 60\% within the emission bandwidth of the scintillators.  Dose monitoring was realized with a computer connected to the photon counter. Miniaturization thus takes benefit of the better performances of narrow silica fibers and single-pixel photon counters regarding plastic fibers and the multipixel photodiodes used with plastic fibers, respectively. 

\begin{figure}[htbp]
\centering\includegraphics[scale=1.19]{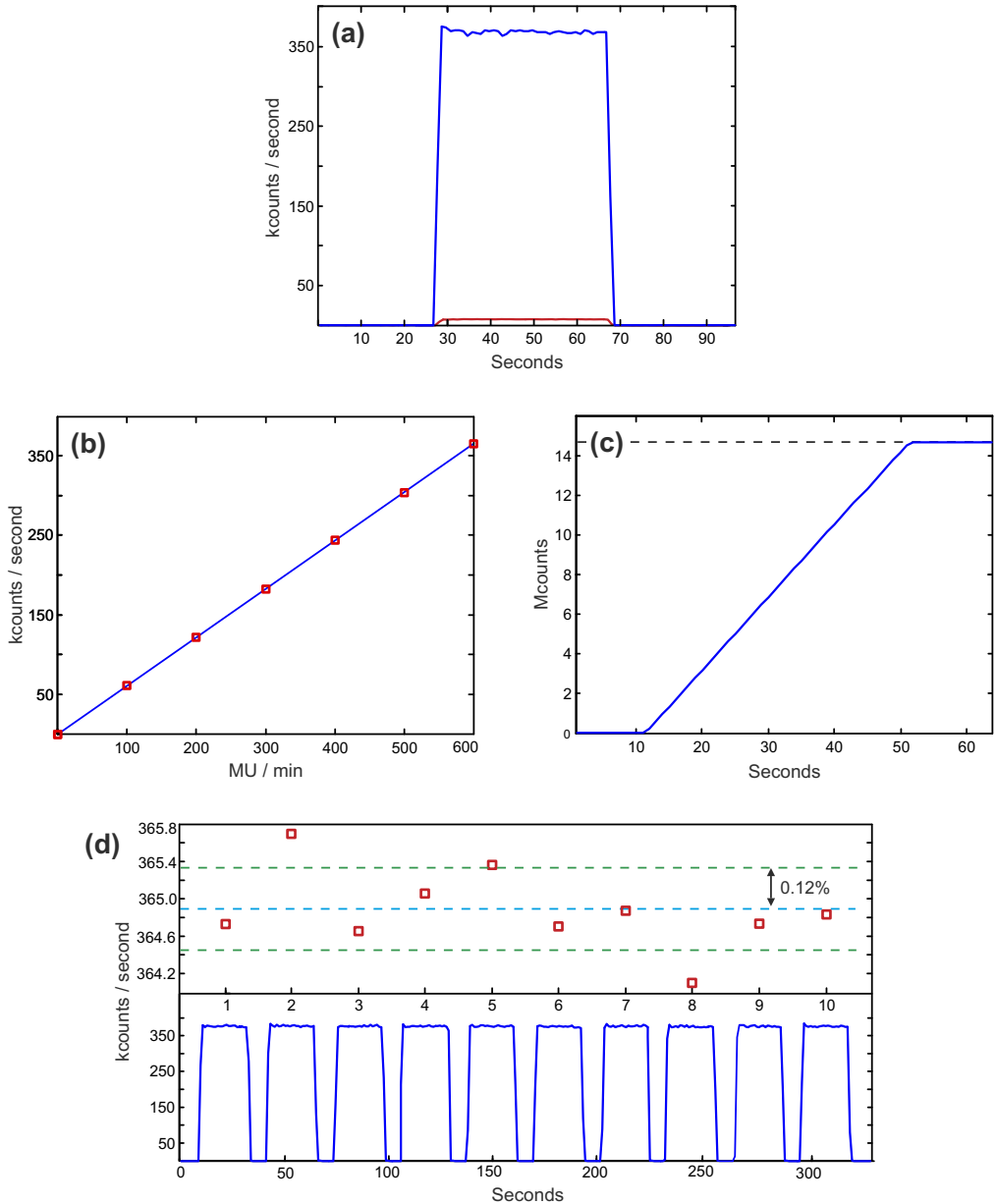}
\caption{(a) Detected optical signal (in photons/s) over time for an exposure at a beam energy of 6 MV and at dose of 300 MU with rate of 600 MU/min. (b) Detected optical signal as a function of the dose rate. (c) Cumulated dose over time deduced by integrating the detected optical signal of (a) (40 seconds exposure at 600 MU/min). (d) Detected optical signal for 10 exposures of 22 seconds at 600 MU/min . Downer panel: detected signal over time. Upper panel: Squares: average signal at each acquisition, dashed lines: Average signal over the 10 exposures and standard deviation (0.12 \% of the average signal).}
\end{figure}

\subsection{Results}

The response of a miniaturized fiber dosimeter was determined for 400 MU (Monitory Units) at a dose rate of 600 MU/min with a field size of $5\times 5$ cm$^2$ and a normal X-ray photon energy of 6 MV. In this investigation, 1 MU is equal to an absorbed dose of 1 cGy under standard reference conditions. The integration time of the photon counter has been set to one second, which is compatible with the real-time monitoring of the dose. Figure 5(a) shows the detection signal upon irradiation. At the beginning and at the end of the acquisition, the irradiator was off to estimate a background (dark) noise level of the photon counter smaller than 15 counts/s. When the radiation beam is activated, a signal larger than 360 kcounts/s was measured with a signal-to-noise ratio of 190. Although such a signal-to-noise level is high enough to perform accurate dose monitoring, the observed signal fluctuations remain 2.98 fold higher than the intrinsic detection noise of the photon counter at this photon flux. Such a discrepancy may be explained either by high frequency fluctuations in the irradiation beam or by an electric perturbation of the optical detector when the irradiator is switched on. Higher signal-to-noise ratio can be obtained by increasing the integration time of the photodetector to a few seconds (two or three seconds to ensure real-time acquisition). Figure 5(a) also shows a Cerenkov signal (measured with a blank fiber) of 7800 counts/s, that is 44 fold smaller than the luminescence signal from the scintillators. Even small, Cerenkov effect within the fiber cannot be neglected in the probe detection process. It can however be easily filtered out in a parallel paired-fiber configuration \cite{wells:ijrobp94,lee:nimprs07,yoo:ox13}. In this case, the use of narrow optical fibers may solve problems of probe compactness encountered with plastic fibers in intracorporeal dosimetry. The resulting two-fiber dosimeters do indeed remain at least two times narrower than the single-fiber probes developed so far \cite{zhuang:ox16,deandreas:ieee17,archer:sr17}. Complex signal processing aimed at removing Cerenkov signal in a single-fiber dosimeter architecture \cite{frelin:mp05,justus:ao04,clift:pmb02} could thus be avoided. 

Fig. 5(b) reports the detected optical signal as a function of the dose rate. This result reveals a linear response of the dosimeter with respect to the radiation power.  The accumulated optical power over time is directly related to the radiation dose applied to the probe. As seen in Fig. 5(c), the time-integrated optical signal shows a constant enhancement of the dose during the probe irradiation, which agrees with the constant dose rate (600 MU/min) applied to the probe. 

The repeatability of the dose measurement with our fiber dosimeter is confirmed in Fig. 5(d). The repeatability was tested over ten exposures of 22 seconds at a dose rate of 600 MU/min with a radiation energy of 6 MV. The detected signal reveals ten plateaus of almost constant signal described by an average value of 364.88 kcounts/s and a relative deviation of 0.12 \% of the average value. 

\section{The benefit of miniaturization: ultra-low footprint fiber probes}

Beyond biocompatibility, \textit{in vivo} X-ray dosimeters must be of negligible footprint for the patient and the therapeutic process, i.e., ultra-compact and invisible to the therapeutic process. It has been reported by Papanikolaou et al that a 5\% change in dose may result in a 10 \% to 20 \% change in tumor control probability at a TCP of 50 \%, as well as in a 20 \% to 30 \% impact on complication rates in normal tissues \cite{papanikolaou:04}. Although these results refer to changes caused by homogeneous dose distributions covering the whole tumor, they stress the need to develop detection tools with a minimum absorption of the incoming radiations, especially in the treatment of small tumors.
 
Fig. 6 reports an estimation of the radiation absorption by our fiber dosimeter. To this end, a model of the probe has been elaborated from the optical images of the structure (see Figs. 4(a) and (b)) and the X-ray mass-attenuation coefficients of its constitutive materials \cite{nist2}. The geometry of our probe model is that of a tapered silica cylinder with a diameter of 125 $\mu$m. The taper angle is 12.4$^{\circ}$ consistent with the angle of the fabricated fiber tip of Fig. 4(a). The cone and the very end of the cylinder are covered with an axis symmetrical layer of gadolimium oxysulfide whose thickness distribution leads to the probe shape of Fig. 4(b). The probe is considered being irradiated from the side, leading to an attenuation area which does not exceed 0.153 mm$^2$.

\begin{figure}[htbp]
\centering\includegraphics[scale=1.48]{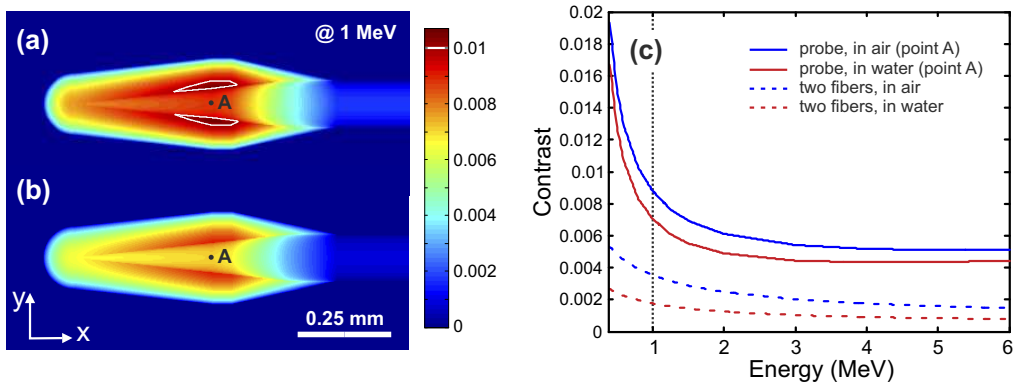}
\caption{(a) and (b) Mappings of the probe-induced intensity contrast of an incoming radiation beam (a) in air and (b) in water. The radiations impinge from the probe side.  (c) Solid lines: spectra of the beam intensity contrast calculated at a single point of the fiber probe positioned in water (red curve) and in air (blue curve). This point is identified with the letter A in (a) and (b). The beam energy ranges from 0.4 MV to 6 MV. Dashed lines: spectra of the maximum beam intensity contrast induced by a couple of narrow 125-micron outer diameter optical fibers.} 
\end{figure}

Fig. 6(a) and (b) show mappings of the probe-induced attenuation of a monochromatic radiation at 1MeV, both in air (Fig. 6(a))  and in water (Fig. 6(b)). These two figures represent intensity contrast after the probe, defined as $T(x,y)=(I_0(x,y)-I(x,y))/I_0(x,y)$ where $I(x,y)$ and $I_0(x,y)$ are the 2D radiation intensity distributions with and without the fiber probe, respectively. In air, the probe-induced intensity contrast peaks at 1.08 \% and shows values exceeding 1\% over an area of only 0.005 mm$^2$ (the total attenuation area is 0.153 mm$^2$).  The narrow optical fiber alone leads to an intensity contrast lower than 0.18\%.  In a water-equivalent (biological) medium, the maximum attenuation of our fiber probe and the fiber alone fall down to 0.9 \% and 0.09 \%, respectively.  Therefore, our fiber probe can be considered of negligible footprint at 1 MeV, both in air and in a biological environment. 

Medical irradiators being polychromatic sources, we also evaluated the probe-induced intensity contrast over a spectral bandwidth ranging from 0.4 MeV to 6 MeV. Fig. 6(c) reports spectra in air and in water at a single point A represented in Fig 6(a) and 6(b). We see that the probe-induced radiation decay remains under 1 \% at energies larger than 0.84 MeV and 0.67 MeV in air and in water, respectively. We also showed maximum beam attenuation through a couple of parallel fibers (configuration used for filtering Cerenkov signal). In that case, maximum intensity contrast of 0.53 \% in air and 0.26 \% in water is predicted at 0.4 MeV. 

These results show that the provided miniaturization opportunity leads to fiber probes of negligible footprint, even with non water-equivalent scintillators and fibers. This greatly relaxes exigencies in the probe design.

\section{Conclusion}

We demonstrated a miniaturized fiber probe for locally measuring in real-time the dose of medical radiations (photons and possibly charge particles). Such a probe relies on the concept of a leaky-wave optical antenna developed at the end of a narrow 125-micron outer diameter optical fiber. The antenna concept is used as a key-connection between the fiber and the scintillators aimed at enhancing the in-fiber coupling of the radiation-triggered luminescence. Miniaturization also takes benefit of the low propagation loss of narrow silica fibers and high detection efficiency of single-pixel photon counters.

The advantage of miniaturization is threefold for the dosimetry of medical radiations. First, the scintillating material being spread along the fiber axis in the form of a thin film (of micrometer thickness), probe-induced radiation decay remains at a very low level even with non water-equivalent scintillators and fibers. Second, the scintillating area (of maximum absorption) is only a tenth of a square millimeter. Finally, the Cerenkov signal could be filtered out in a simple and low-cost parallel paired fiber configuration which remains ultracompact and compatible with intracorporeal dosimetry for radiotherapy applications. Miniaturization may thus provide a new strategy to reach \textit{in vivo} fiber dosimeters of negligible footprint for the patient and the therapeutic process, even with non water-equivalent scintillators and fibers. 

Our concept of a miniaturized fiber dosimeter opens new perspectives for radiation dose monitoring and may enable unprecedented control in up-to-date and future radiation therapies. As an example, ultracompact and ultralow footprint multiprobe fiber dosimeters may emerge from a parallel implementation of our fiber probe onto a bundle of fibers of different lengths. The opportunity of engineering one probe per fiber avoids cross-talk between parallel acquisition channels as compared to existing multi-probe devices sharing the same fiber (due to compactness limitations inherent to the use of broad plastic fibers) \cite{agostino:patent17}. Such multiprobe dosimeters could be of crucial importance in the \textit{in vivo} localization and metrology of sharp dose gradients. 

Our fiber probes fulfill the increasing requirement of accuracy in medical dose delivery and are compatible with the current image guidance techniques for target positioning (such as MRI). Owing to their ultra-small detection cell, they are well adapted to all current radiation therapies as well as to the newborn microbeam radiation therapy. 


\section*{Acknowledgments}
The authors are indebted to the company SEDI-ATI for helpful discussions. This study was conducted in close connection with this company in the purpose of probe optimization and encapsulation. This work was funded by the CNRS DEFI Instrumentation aux limites 2018, the french agency of research (program ANR-18-CE42-0016), the Region Bourgogne Franche-Comte and the EIPHI Graduate School (ANR-17-EURE-0002).


\end{document}